\documentclass{IEEEtran}
\usepackage{cite}
\usepackage{amsmath,amssymb,amsfonts}
\usepackage{algorithm}
\usepackage{algorithmic}
\usepackage{graphicx}
\usepackage{textcomp}

\usepackage{xcolor}
\def\BibTeX{{\rm B\kern-.05em{\sc i\kern-.025em b}\kern-.08em
    T\kern-.1667em\lower.7ex\hbox{E}\kern-.125emX}}

\floatname{algorithm}{Algorithm}

\newtheorem{theorem}{Thoerem}
\newtheorem{lemma}{Lemma}
\newtheorem{assumption}{Assumption}
\newtheorem{remark}{Remark}

\newcommand{\diff}{\mathrm{d}}
\newcommand{\e}{\mathrm{e}}
\newcommand{\R}{\mathbb{R}}
\newcommand{\Z}{\mathbb{Z}}
\newcommand{\T}{\mathsf{T}}
\renewcommand{\tilde}{\widetilde}
\renewcommand{\hat}{\widehat}

\begin{document}

\title{Linearization-Based Quantized Stabilization \\ of Nonlinear Systems Under DoS Attacks}
\author{Rui~Kato,
  Ahmet~Cetinkaya,~\IEEEmembership{Member,~IEEE,}
  and~Hideaki~Ishii,~\IEEEmembership{Senior~Member,~IEEE}
  \thanks{Manuscript received April 7, 2020. This work was supported in part by the JST CREST Grant No.~JPMJCRl5K3, in part by JSPS under Grant-in-Aid for Scientific Research Grant
  No.~18H01460, and in part by JST ERATO HASUO Metamathematics for Systems Design Project (No.~JPMJER1603).}
  \thanks{R. Kato and H. Ishii are with the Department of Computer Science, Tokyo Instutite of Technology, Yokohama 226-8502, Japan (e-mail: kato@sc.dis.titech.ac.jp; ishii@c.titech.ac.jp).}
  \thanks{A. Cetinkaya is with with Information Systems Architecture Science Research Division, National Institute of Informatics, Tokyo 101-8430, Japan (e-mail: cetinkaya@nii.ac.jp).}
}

\maketitle

\begin{abstract} 
Motivated by recent security issues in cyber-physical systems, this technical note studies the stabilization problem of networked control systems under Denial-of-Service (DoS) attacks.
In particular, we consider to stabilize a nonlinear system with limited data rate via linearization.
We employ a deterministic DoS attack model constrained in terms of attacks' frequency and duration, allowing us to cover a large class of potential attacks.
To achieve asymptotic stabilization, we propose a resilient dynamic quantizer in the sense that it does not saturate in the presence of packet losses caused by DoS attacks.
A sufficient condition for stability is derived by restricting the average DoS frequency and duration.
In addition, because of the locality of linearization, we explicitly investigate an estimate of the region of attraction, which can be expected to be reduced depending on the strength of DoS attacks.
A simulation example is presented for demonstration of our results.
\end{abstract}


\section{Introduction}
\label{sec:introduction}

Networked control systems have been widely studied over the past several decades \cite{Bemporad2010}.
In recent years, cyber security of such systems has attracted much attention as the communication channels are exposed to malicious attackers; see, e.g., \cite{Cardenas2008} and \cite{Pasqualetti-Dorfler-Bullo:CSM2015} for an overview.
It has become clear that cyber attacks to control systems may induce critical incidents in the real world, resulting in, e.g., physical damages in equipments and financial losses.
The authors of \cite{Amin2009} classified cyber attacks on control systems into \emph{deception attacks}, which are conducted by changing the contents of packet data, and \emph{Denial-of-Service (DoS) attacks}, which refer to communication interruptions including jamming attacks.
DoS attacks are particularly critical as it is easier to launch than deception attacks as mentioned in \cite{Teixeira2015}.
For this reason, we examine the effects of DoS attacks in this paper.

Since it is not rational to assume that malicious attacks follow a certain probability distribution, we treat DoS attacks in a deterministic manner rather than a stochastic one; see the survey paper \cite{Cetinkaya2019a} for more detailed discussions on various DoS attack models.
A characterization of deterministic DoS attacks in terms of average frequency and duration was introduced by \cite{DePersis-Tesi:TAC2015}, and is also used in this paper.
In that paper, allowable DoS frequency and duration to guarantee input-to-state stability of linear systems were obtained.
These conditions were made less conservative in \cite{Feng-Tesi:automatica2017} by using a predictor that estimates interrupted measurements.
On the other hand, global stability of nonlinear systems under DoS attacks was investigated in \cite{DePersis-Tesi:sysconle2016}.
In contrast, the paper \cite{Cetinkaya-Ishii-Hayakawa:TAC2017} provided a comprehensive treatment of both malicious and non-malicious packet losses.
A switched system framework was also studied in \cite{Cetinkaya-Ishii-Hayakawa:TAC2019}.

On the other hand, data rate limitation of communication channels is one of the important issues in networked control systems \cite{Ishii2002}.
In this context, information to be exchanged over communication networks must be quantized.
Many researchers have explored a range of quantized control problems from various perspectives; see, e.g., \cite{Nair2007} and the references therein.
For considering asymptotic stabilization under the required data rate, we employ time-varying quantizers with the zooming-in and zooming-out capabilities proposed by \cite{Liberzon-Hespanha:TAC2005}.
However, packet losses may induce saturation of the dynamic quantizer, since its quantization region becomes small as time passes.
To avoid such situations, we propose the resilient design that expands the quantization region depending on the occurrence of DoS attacks.
Recently, observer-based quantized control under DoS attacks was considered in \cite{Wakaiki2019}.
In \cite{Feng-Cetinkaya-Ishii-Tesi-DePersis:TAC2020}, the trade-off between the minimum data rate for stabilization and the tolerable level of DoS attacks was revealed.
Furthermore, the minimum data rate problem in the presence of probabilistic packet losses has been addressed in \cite{You2011} and \cite{Minero2013}.
These results are applicable to linear systems but not to nonlinear systems.
In this paper, we consider quantized control of nonlinear systems via linearization as studied in \cite{Hu1999}.

Though linearization-based control is a typical method in practice, the effects of DoS attacks have not been much explored in the literature.
It is of particular interest in the context of DoS attacks, since they may bring critical issues when communication is interrupted.
Indeed, if the state leaves the region of attraction due to DoS attacks, then it will not converge to the equilibrium point even after the communication is restored.
In \cite{Kato2019ACC}, a linearization approach was analyzed and an estimate of the region of attraction under DoS attacks was derived.
This paper provides an extension of the framework presented there to take quantization effects into account.

The subsequent sections are organized as follows.
In Section~\ref{sec:problem}, we describe the problem setting and the DoS attack model used in this paper.
The encoding/decoding scheme and the proposed resilient dynamic quantizer are introduced in Section~\ref{sec:quantized_control}.
The main results of this paper are presented in Section~\ref{sec:stability_analysis}, where a sufficient condition for stability and an initial condition to guarantee the convergence of state trajectories are derived.
In Section~\ref{sec:simulation}, we present a simulation example.
Finally, we conclude the paper in Section~\ref{sec:conclusion}.
The preliminary version of this paper appeared as \cite{RK-AC-HI:IFAC2020}.
The current paper contains full proofs of the results.

Throughout this paper, we employ the following notation.
The sets of nonnegative reals and nonnegative integers are denoted by $\R_+$ and $\Z_+$, respectively.
Given a vector $v$ and a matrix $M$, $\|v\|_\infty$ and $\|M\|_\infty$ respectively denote the $\infty$-norm and the induced $\infty$-norm.
The length of an interval $\mathcal{I}$ is denoted by $|\mathcal{I}|$.

\section{Problem Formulation}
\label{sec:problem}

In this section, we describe the problem setting of networked control and the DoS attack model characterized by their frequency and duration.

\subsection{Nonlinear Networked Control System}

\begin{figure}
  \centering
  \includegraphics[width=0.7\columnwidth]{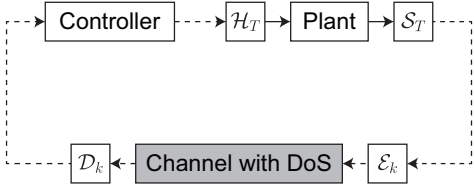}
  \caption{Networked control system under DoS attacks}
  \label{fig:ncs}
\end{figure}

Consider the nonlinear networked control system depicted in Fig.~\ref{fig:ncs}, where a communication channel is inserted between the sensor and the controller.
Here, the plant to be controlled is described by
\begin{align}
  \dot{x}(t) = f(x(t),u(t)), \quad t \ge 0,
  \label{eq:plant}
\end{align}
where $x(t) \in \R^n$ is the state and $u(t) \in \R^m$ is the control input at time $t$.
The initial state is given by $x(0) = x_0 \in \R^n$.
Assume that $f \colon \R^n \times \R^m \to \R^n$ is continuously differentiable and that the system \eqref{eq:plant} has an equilibrium point at the origin, i.e., $f(0,0) = 0$.
Then, we impose the following assumption.

\begin{assumption}
  The function $f$ in \eqref{eq:plant} is Lipschitz in a certain region $\mathcal{D} := \{x \in \R^n : \|x\|_\infty < \varrho\}$ for any input $u \in \R^m$, where $\varrho > 0$ is some positive number.
  That is, there is a constant $L \ge 0$ satisfying $\|f(y,u) - f(z,u)\|_\infty \le L \|y - z\|_\infty$ for all $y,z \in \mathcal{D}$ and $u \in \R^m$.
  \label{assump:Lipschitz}
\end{assumption}

Letting $T > 0$ be a fixed sampling period, we denote by $t_k := kT$, $k \in \Z_+$, the sampling instants.
The ideal sampler $\mathcal{S}_T$ measures the state at each sampling time.
The sampled state is then transformed by the encoder $\mathcal{E}_k$ into a certain symbol to be sent through the communication channel.
At the controller side, the decoder $\mathcal{D}_k$ produces the quantized state after receiving the packet as explained in the next section.
During the sampling/transmission intervals, the control input is kept constant by the zeroth-order hold $\mathcal{H}_T$.

For given vectors $\bar{x} \in \R^n$ and $\bar{u} \in \R^m$, let $\phi(t,\bar{x},\bar{u})$ be the solution to \eqref{eq:plant} for $t \in [0,T]$ with the initial state $x_0 = \bar{x}$ and the constant input $u(t) \equiv \bar{u}$.
Then, we define $\phi_T(\bar{x},\bar{u}) := \phi(T,\bar{x},\bar{u})$.
Furthermore, for ease of presentation, we write the sampled value $x(t_k)$ as $x_k$ for each $k \in \Z_+$, and the same notation is used for other variables as well.

If a DoS attack is active at a sampling time, then the packet transmission at that instant fails.
In this case, the control input is set to zero until the next packet reaches the controller side.
Let $\theta_k \in \{0,1\}$ be the indicator that stands for the absence or presence of packet losses.
If a packet loss occurs at time $t_k$, we set $\theta_k = 1$, and otherwise $\theta_k = 0$.
Then, the control input applied to the plant \eqref{eq:plant} is given as follows:
\begin{align}
  u(t) = (1 - \theta_k) Kq_k, \quad t \in [t_k,t_{k+1}), \quad k \in \Z_+,
  \label{eq:control}
\end{align}
where $K \in \R^{m \times n}$ is a feedback gain matrix, the choice of which is given later.
Moreover, $q_k \in \R^n$ denotes the quantized value of the sampled state $x_k$.

\subsection{Data Rate Limitation}

Since we consider a communication channel whose data rate is limited, the information that the packet can contain is taken from a finite set.
Let $\mathcal{M} := \{0,1,\ldots,M^n - 1\}$ be the set of integers that can be sent by communication at each transformation, where $M$ is a positive integer expressing the number of the quantization levels in one coordinate of $\R^n$.
In this case, the data rate of the channel is denoted by $R := n \log_2(M)/T$ bits per unit of time.
Defining $\Lambda := \e^{LT}$, in what follows, we make the assumption below.

\begin{assumption}
  The number of the quantization levels $M$ satisfies $M > \Lambda$.
  \label{assump:data_rate}
\end{assumption}

\begin{remark} \label{rem:data_rate}
  The above condition can be found in \cite{Liberzon-Hespanha:TAC2005}, and it is sufficient to stabilize the nonlinear system \eqref{eq:plant} if there is no packet loss.
  Thus, the conservativeness of the data rate condition is the same as that in \cite{Liberzon-Hespanha:TAC2005}, although DoS attacks are considered.
  Note that, for linear systems, one can reduce the data rate condition using a certain coordinate transformation as considered in \cite{Wakaiki2019,Feng-Cetinkaya-Ishii-Tesi-DePersis:TAC2020}.
  However, for nonlinear systems, it is difficult to find such a transformation.
  Although local asymptotic stability can be preserved under a data rate which is arbitrarily close to the minimum data rate for the linearized system \cite{Nair2004}, it is not practically enough from the viewpoint of, e.g., the region of attraction.
  Because we quantitatively explore the region of attraction in the subsequent section, the above assumption on the data rate is employed.
\end{remark}

\subsection{Averagely Constrained DoS Attacks}

Here, we introduce a deterministic class of DoS attacks.
For $i \in \Z_+$, let $a_i \ge 0$ and $\tau_i \ge 0$ denote the launching time and the length of the $i$th DoS attack, respectively.
Notice that when $\tau_i = 0$, the attack is impulsive, and thus, it has no length.
We then define the collection of DoS attack intervals by
\begin{align*}
  \mathcal{A}(t) := \bigcup_{i \in \Z_+} [a_i,a_i + \tau_i] \cap [0,t].
\end{align*}
Furthermore, we denote by $N(t)$ the number of DoS attacks for which the starting time is inside the interval $[0,t]$.
Following the work of \cite{DePersis-Tesi:TAC2015}, we characterize DoS attacks in terms of their frequency and duration.

\begin{assumption}[DoS frequency]
  There exist constants $\kappa_F \ge 0$ and $\rho_F \in [0,\infty)$ such that
  \begin{align*}
    N(t) \le \kappa_F + \rho_F t, \quad t \ge 0.
  \end{align*}
  \label{assump:frequency}
\end{assumption}

\begin{assumption}[DoS duration]
  There exist constants $\kappa_D \ge 0$ and $\rho_D \in [0,1)$ such that
  \begin{align*}
    |\mathcal{A}(t)| \le \kappa_D + \rho_D t, \quad t \ge 0.
  \end{align*}
  \label{assump:duration}
\end{assumption}

In the above assumptions, the constants $\rho_F$ and $\rho_D$ represent the allowable average frequencies and durations of DoS attacks.
On the other hand, the constants $\kappa_F$ and $\kappa_D$ indicate the initial energy to launch attacks.
In this framework, an attacker does not need to follow certain attack strategies such as periodic attacks.
Note that an attacker can launch frequent but short DoS attacks to cause packet losses at all transmission times.
Such situations may occur when $\rho_F \ge 1/T$ is allowed, under which DoS attacks can be sufficiently frequent compared with the transmission period.
This implies that periodic communications are vulnerable as the transmission time instants are available for attackers.
To make the communication more secure, randomized transmission protocols are proposed by \cite{Cetinkaya-Kikuchi-Hayakawa-Ishii:automatica2020} in the context of multi-agent consensus problems.

\begin{remark} \label{rem:dos}
  In \cite{DePersis-Tesi:TAC2015,Feng-Tesi:automatica2017}, more restrictive class of DoS attacks is considered.
  There, the frequencies and the durations of DoS attacks are constrained for any time intervals $[\tau,t]$ with $\tau \le t$ rather than $[0,t]$.
  Note that such assumptions are required to guarantee input-to-state stability with respect to disturbances \cite{DePersis-Tesi:TAC2015} or to construct a state predictor \cite{Feng-Tesi:automatica2017}.
  In particular, the DoS model considered in \cite{DePersis-Tesi:TAC2015,Feng-Tesi:automatica2017} has an upper bound on the consecutive packet losses.
  In contrast, we do not assume consecutive packet losses to be bounded.
  We also note that the DoS parameters are determined depending on the attacker's resource.
  As the attacker's power is time-varying, these parameters can be time dependent in general.
  However, the control parameters are fixed in this paper, and hence, we only consider the constant DoS parameters.
  If one employs adaptive or switching control strategies whose parameters are changed depending on the attack level in real time, then there is an advantage to estimate the DoS parameters on-line.
\end{remark}

\section{Quantized Control via Linearization}
\label{sec:quantized_control}

In this section, we consider to stabilize the nonlinear system \eqref{eq:plant} via linearization.
First, we explore the inter-sample behavior and the vanishing perturbation property of the remainder term of linearization.
Then, the encoding and decoding procedures are explained, followed by proposing a resilient dynamic quantizer design.

\subsection{Linearization Analysis}

Linearization of \eqref{eq:plant} around the origin yields
\begin{align}
  \dot{x}(t) = Ax(t) + Bu(t) + g(x(t),u(t)),
  \label{eq:linearization}
\end{align}
where
\begin{align*}
  A := \left. \frac{\partial f(x,u)}{\partial x} \right|_{x=0,u=0}, \quad B := \left. \frac{\partial f(x,u)}{\partial u} \right|_{x=0,u=0},
\end{align*}
and $g(x,u) := f(x,u) - Ax - Bu$ is the remainder term of the linear approximation.
Assume that $A$ is unstable and that the pair $(A,B)$ is stabilizable.

Then, we discretize the continuous-time system \eqref{eq:linearization} with sampling period $T$ to obtain
\begin{align}
  x_{k+1} = \tilde{A} x_k + \tilde{B} u_k + \tilde{g}(x_k,u_k),
  \label{eq:discretization}
\end{align}
where $\tilde{A} := \e^{AT}$, $\tilde{B} := \int_0^T \e^{As} \,\diff s B$, and
\begin{align*}
  \tilde{g}(x_k,u_k) := \int_0^T \e^{A (T - s)} g(\phi(s,x_k,u_k),u_k) \,\diff s.
\end{align*}
Here, we suppose that the sampling period is nonpathological, and hence, $(\tilde{A},\tilde{B})$ is stabilizable.
We now choose the controller gain $K$ in \eqref{eq:control} such that $\tilde{A} + \tilde{B} K$ is Schur stable.
By this choice, the origin $x = 0$ is locally asymptotically stable for \eqref{eq:plant} in the absence of DoS attacks.
Note that global stability is not guaranteed due to linearization, which is important in the context of networked control under DoS attacks.

Whereas \cite{Wakaiki2019} considers discrete-time systems, we employ the sampled-data setting as bounds on the inter-sample behavior are required to analyze the plant nonlinearity.
We now define $c_0 := [1 + T (\|BK\|_\infty + \|K\|_\infty)] \e^{T (\|A\|_\infty + 1)}$ and $c_1 := \e^{T (\|A\|_\infty + 1)}$.
The following lemma is useful to examine bounds on the nonlinear term in \eqref{eq:discretization}.

\begin{lemma}
  For any $\bar{x} \in \R^n$, consider the solution $\phi(t,\bar{x},\bar{u})$ to \eqref{eq:plant} with $\bar{u} = (1 - \theta) K \bar{x}$, where $\theta \in \{0,1\}$.
  Then, there exists a constant $d > 0$ such that $\|\bar{x}\|_\infty < d$ implies for all $t \in [0,T)$,
  \begin{align*}
    \|\phi(t,\bar{x},\bar{u})\|_\infty \le
    \begin{cases}
      c_0 \|\bar{x}\|_\infty & \text{if $\theta = 0$}, \\
      c_1 \|\bar{x}\|_\infty & \text{if $\theta = 1$}.
    \end{cases}
  \end{align*}
  \label{lem:inter-sample}
\end{lemma}

\begin{IEEEproof}
  See Appendix~\ref{app:inter-sample}.
\end{IEEEproof}

To explore local stability of the origin, we need bounds on the remainder term of linearization.
Given $\gamma > 0$, we define $\gamma_0 := (c_0 + \|K\|_\infty) \gamma T \e^{T \|A\|_\infty}$ and $\gamma_1 := c_1 \gamma T \e^{T \|A\|_\infty}$.
In the following lemma, we give the region inside which the growth of the effects of the plant nonlinearity is characterized in terms of the state norm.

\begin{lemma}
  For any $\bar{x} \in \R^n$, consider the nonlinear function $\tilde{g}(\bar{x},\bar{u})$ in \eqref{eq:discretization} with $\bar{u} = (1 - \theta) K \bar{x}$, where $\theta \in \{0,1\}$.
  Then, for every $\gamma > 0$, there exists a constant $\delta \in (0,d]$ such that $\|\bar{x}\|_\infty < \delta$ implies
  \begin{align*}
    \|\tilde{g}(\bar{x},\bar{u})\|_\infty \le
    \begin{cases}
      \gamma_0 \|\bar{x}\|_\infty & \text{if $\theta = 0$}, \\
      \gamma_1 \|\bar{x}\|_\infty & \text{if $\theta = 1$},
    \end{cases}
  \end{align*}
  where $d$ is as in Lemma~\ref{lem:inter-sample}.
  \label{lem:nonlinearity}
\end{lemma}

\begin{IEEEproof}
  See Appendix~\ref{app:nonlinearity}.
\end{IEEEproof}

\subsection{Encoding/Decoding Scheme}

Due to the limited data rate, we consider a finite number of partitions of the quantization region.
In this subsection, we state the encoding/decoding scheme of the dynamic quantizer following \cite{Liberzon-Hespanha:TAC2005}.
We consider the encoder and the decoder which have two time-dependent variables: the center of the quantization region and the radius of the quantization range.
We denote these variables with the symbols $\xi_k \in \R^n$ and $E_k \ge 0$, respectively.
Now, we define the quantization region at time $t_k$ as follows:
\begin{align*}
  \mathcal{Q}(\xi_k,E_k) := \{x \in \R^n : \|x - \xi_k\|_\infty \le E_k\}.
\end{align*}
This is a hypercube which has the edges of length $2E_k$ and is centered at $\xi_k$, and this region must be the same in both the encoder and the decoder at each time.
Since the initial state is not known exactly in general, we set $\xi_0 = 0$.
For $E_0$, we make the following assumption, under which the encoder and the decoder know how far the state is from the origin.

\begin{assumption}
  We set $E_0 \ge 0$ such that the initial state $x_0$ of \eqref{eq:plant} satisfies $\|x_0\|_\infty \le E_0$.
  \label{assump:initial_state}
\end{assumption}

To avoid saturation of the quantizer, $\xi_k$ and $E_k$ are adjusted based on the reachable set of state trajectories.
In this paper, we assume that an acknowledgement signal or the value of $\theta_k$ is exchanged between the encoder and decoder and that this signal is not subject to DoS attacks similarly to \cite{Wakaiki2019} and \cite{Feng-Cetinkaya-Ishii-Tesi-DePersis:TAC2020}.
In practice, this assumption is unrealistic.
However, one can estimate the occurrence of packet losses from the behavior of the state without acknowledgements as considered in \cite{Ishii_automatica09}.


\begin{algorithm}[!tb]
  \caption{Encoding process}
  \label{alg:encoding}
  \begin{algorithmic}
    \REQUIRE Sampled state measurement $x_k \in \mathcal{Q}(\xi_k,E_k)$.
    \ENSURE Encoded symbol $i \in \mathcal{M}$.
    \STATE The quantization region $\mathcal{Q}(\xi_k,E_k)$ is partitioned into the $M^n$ equal boxes with the same dimension, each of which is indexed by an integer in $\mathcal{M}$.
    \FOR{$k \in \Z_+$}
      \STATE Encode $x_k$ into the symbol $i$ associated with the partition in which $x_k$ lies.
      \STATE Send the symbol $i$ to the decoder.
      \STATE Receive an acknowledgement $\theta_k \in \{0,1\}$ from the decoder.
      \STATE Update $\xi_k$ and $E_k$ based on the value of $\theta_k$ by the rules \eqref{eq:xi} and \eqref{eq:E}.
    \ENDFOR
  \end{algorithmic}
\end{algorithm}

\begin{algorithm}[!tb]
  \caption{Decoding process}
  \label{alg:decoding}
  \begin{algorithmic}
    \REQUIRE Encoded symbol $i \in \mathcal{M}$.
    \ENSURE Quantized state measurement $q_k \in \R^n$.
    \STATE The decoder knows which symbol $i \in \mathcal{M}$ corresponds to which partition of $\mathcal{Q}(\xi_k,E_k)$.
    \FOR{$k \in \Z_+$}
      \IF{if the decoder receives the packet at time $t_k$}
        \STATE Set $q_k$ as the center $\xi_k$ of the partition associated with the received symbol $i$.
        \STATE Send the acknowledgement $\theta_k = 0$ to the encoder.
      \ELSE
        \STATE Set $q_k$ to zero.
        \STATE Send the acknowledgement $\theta_k = 1$ to the encoder.
      \ENDIF
      \STATE Update $\xi_k$ and $E_k$ based on the value of $\theta_k$ by the rules \eqref{eq:xi} and \eqref{eq:E}.
    \ENDFOR
  \end{algorithmic}
\end{algorithm}

The encoding and decoding processes are described in Algorithms~\ref{alg:encoding} and \ref{alg:decoding}, respectively.
If we know which partitioned box the state lies in, then the reachable set at the next sampling instant can be estimated so that it becomes smaller than the current quantization region, resulting in the zooming-in process.
However, if the packet loss occurs at time $t_k$, we know only that the state $x_k$ is inside $\mathcal{Q}(\xi_k,E_k)$.
Hence, one needs to expand the quantization region to capture the state $x_{k+1}$ at the next sampling time $t_{k+1}$, leading to the zooming-out process.
In the next subsection, we explain how the quantizer is updated depending on the value of $\theta_k$ while the effects of DoS attacks are taken into account.

\subsection{Resilient Dynamic Quantizer Design}

Suppose now that the sampled state $x_k$ lies in the quantization region $\mathcal{Q}(\xi_k,E_k)$, which is equivalent to $\|x_k - \xi_k\|_\infty \le E_k$.
Recall from the quantization procedure mentioned above, $q_k$ is the center of the partitioned box in which $x_k$ lies.
Thus, we know that the quantization error satisfies
\begin{align}
  \quad \|x_k - q_k\|_\infty \le \frac{1}{M} E_k.
  \label{eq:boundary_condition}
\end{align}
To avoid saturation of the quantizer, i.e., to ensure that the state never goes outside the quantization region, both the encoder and decoder need to calculate $\xi_{k+1}$ and $E_{k+1}$ so that the following inequality holds:
\begin{align}
  \|x_{k+1} - \xi_{k+1}\|_\infty \le E_{k+1},
  \label{eq:quantizer_condition}
\end{align}
which is equivalent to $x_{k+1} \in \mathcal{Q}(\xi_{k+1},E_{k+1})$.

To do so, we propose the following update rules:
At each sampling time $t_k$, the encoder and decoder generate $\xi_{k+1}$ and $E_{k+1}$ by
\begin{align}
  \xi_{k+1} &:=
  \begin{cases}
    \phi_T(q_k,Kq_k) & \text{if $\theta_k = 0$}, \\
    \phi_T(\xi_k,0)  & \text{if $\theta_k = 1$},
  \end{cases}
  \label{eq:xi} \\
  E_{k+1} &:=
  \begin{cases}
    \dfrac{\Lambda}{M} E_k & \text{if $\theta_k = 0$}, \\
    \Lambda E_k            & \text{if $\theta_k = 1$}.
  \end{cases}
  \label{eq:E}
\end{align}

The zooming-in/out process depends on some variables.
First, $\xi_k$ is updated to trace the state trajectory by estimating the reachable set at time $t_{k+1}$.
This process is conducted by simulating the nonlinear system model.
Since our main focus is local stabilization via linearization-based control, we do not consider the computational complexity.
Second, $E_k$ is updated to cover the uncertainty on the estimate of the reachable set.
Such uncertainty can be known from the Lipschitz property of the system \eqref{eq:plant}, which is assumed in Assumption~\ref{assump:Lipschitz}.
In particular, the quantization level $M$ must large enough such that the trajectory remains in the quantization region.
If there are some uncertainties such as unmodeled dynamics and computation errors, then one can modify the zooming rate in \eqref{eq:E} to avoid the saturation of the quantizer.

The quantizer needs to be capable to expand its quantization range when packet losses occur.
In what follows, we show that the dynamic quantizer with \eqref{eq:xi} and \eqref{eq:E} locally satisfies the condition \eqref{eq:quantizer_condition} at times when both zooming-in and zooming-out occur.

\subsubsection{Zooming-In Process}

We first consider the case where the packet transmission at time $t_k$ is successful, that is, $\theta_k = 0$.
In this case, the quantized state $q_k$ is available for both the encoder and decoder.
Note that from the Lipschitz condition in Assumption~\ref{assump:Lipschitz}, $\|\phi_T(x,u) - \phi_T(y,u)\|_\infty \le \e^{LT} \|x - y\|_\infty$.
Hence, if $x_k,q_k \in \mathcal{D}$, where $\mathcal{D}$ is given in Assumption~\ref{assump:Lipschitz}, then we can see from \eqref{eq:xi} that
\begin{align*}
  \|x_{k+1} - \xi_{k+1}\|_\infty
  &= \|\phi_T(x_k,Kq_k) - \phi_T(q_k,Kq_k)\|_\infty \\
  &\le \Lambda \|x_k - q_k\|_\infty \le \frac{\Lambda}{M} E_k,
\end{align*}
where the last inequality follows from the boundary condition \eqref{eq:boundary_condition}.
Hence, by \eqref{eq:E}, we can guarantee the condition \eqref{eq:quantizer_condition}.
We note that, under Assumption~\ref{assump:data_rate}, the quantization region becomes smaller in the absence of DoS attacks.

\subsubsection{Zooming-Out Process}

We then consider the case where the communication fails at time $t_k$ due to DoS attacks, that is, $\theta_k = 1$.
In this case, the decoder does not know the value of $q_k$ but knows that of $\xi_k$, and thus, the update rule \eqref{eq:xi} can be performed.
Whenever $x_k,q_k \in \mathcal{D}$, we have
\begin{align*}
	\|x_{k+1} - \xi_{k+1}\|_\infty &= \|\phi_T(x_k,0) - \phi_T(\xi_k,0)\|_\infty \\
	                               &\le \Lambda \|x_k - \xi_k\|_\infty \le \Lambda E_k.
\end{align*}
Therefore, the update rules \eqref{eq:xi} and \eqref{eq:E} can be used to ensure that \eqref{eq:quantizer_condition} holds.
Notice that the quantization range becomes larger since $\Lambda > 1$.
This also indicates that DoS attacks induce the expansion of the quantization region.

In \cite{Liberzon-Hespanha:TAC2005}, the zooming-out process is used when the initial state is unknown.
In contrast, our update rule is needed to absorb the effects of DoS attacks.
Moreover, differently from stochastic packet losses, an attacker can launch long DoS attacks to block packet transmissions consecutively.
In our framework, such DoS attacks are constrained by Assumptions~\ref{assump:frequency} and \ref{assump:duration}.

\section{Main Results}
\label{sec:stability_analysis}

In this section, we consider stability analysis of the nonlinear system \eqref{eq:plant} with the control input \eqref{eq:control}.
Furthermore, we provide the initial condition to guarantee the convergence of state trajectories.

\subsection{Characterization of Switched Lyapunov Function}

Various ways to analyze asymptotic stability of switched systems with quantization have been considered such as a switched Lyapunov function approach \cite{Liberzon2014} and a common Lyapunov function approach \cite{Wakaiki2017}.
Differently from the aforementioned papers, we consider both stable and unstable modes.
To handle unstable dynamics, we employ a slightly different technique that captures the system's behavior within the Lyapunov framework.
Furthermore, we deal with nonlinearity of the plant, which affects the increase and decrease rates of a Lyapunov function in a certain region.

Take $\varphi_0 \in (0,1)$ and $\varphi_1 \in (1,\infty)$ to be scalars with which $\varphi_0^{-1/2} (\tilde{A} + \tilde{B} K)$ and $\varphi_1^{-1/2} \tilde{A}$ are Schur stable, respectively.
Then, there exist positive-definite matrices $P_0,P_1 \in \R^{n \times n}$ such that
\begin{align}
  (\tilde{A} + \tilde{B} K)^\T P_0 (\tilde{A} + \tilde{B} K) - \varphi_0 P_0 &\prec 0,
  \label{eq:Lyapunov_inequality1} \\
  \tilde{A}^\T P_1 \tilde{A} - \varphi_1 P_1 &\prec 0.
  \label{eq:Lyapunov_inequality2}
\end{align}
We here note that there always exists a common matrix $P = P_0 = P_1$ if the constant $\varphi_1$ are large enough.
However, more preferable stability condition can be obtained by allowing the use of distinct $P_0$ and $P_1$.
Following the work of \cite{Liberzon2014}, we define for $p \in \{0,1\}$ the positive definite function $W_p \colon \R^n \times \R_+ \to \R_+$ as follows:
\begin{align}
  W_p(\xi,E) := \xi^\T P_p \xi + \eta_p E^2, \quad \xi \in \R^n, \quad E \ge 0,
  \label{eq:Lyapunov_function}
\end{align}
where $\eta_0,\eta_1 > 0$ are sufficiently large numbers.
These functions satisfy the following two properties.
First, there exist $\alpha,\beta > 0$ such that for every $p \in \{0,1\}$,
\begin{align}
  \alpha (\|\xi\|_\infty + E)^2 \le W_p(\xi,E) \le \beta (\|\xi\|_\infty + E)^2.
  \label{eq:alpha_beta}
\end{align}
Second, there exist $\mu_0,\mu_1 \ge 1$ such that
\begin{align}
  W_1(\xi,E) \le \mu_0 W_0(\xi,E), \quad W_0(\xi,E) \le \mu_1 W_1(\xi,E).
  \label{eq:mu}
\end{align}
These properties are not difficult to verify.
For example, to satisfy the first property, we can use
\begin{align*}
  \alpha &= \frac{1}{2} \min_{p \in \{0,1\}} \{\lambda_\mathrm{min}(P_p),\eta_p\}, \\
  \beta &= \max_{p \in \{0,1\}} \{n \lambda_\mathrm{max}(P_p),\eta_p\},
\end{align*}
where $\lambda_\mathrm{min}(\cdot)$ and $\lambda_\mathrm{max}(\cdot)$ represent the smallest and the largest eigenvalues of a matrix, respectively.
Moreover, the following constants can be used for the second property:
\begin{align}
  \mu_0 &= \max \left\{ \frac{\lambda_\mathrm{max}(P_1)}{\lambda_\mathrm{min}(P_0)},\frac{\eta_1}{\eta_0} \right\}, \label{eq:mu0} \\
  \mu_1 &= \max \left\{ \frac{\lambda_\mathrm{max}(P_0)}{\lambda_\mathrm{min}(P_1)},\frac{\eta_0}{\eta_1} \right\}. \label{eq:mu1}
\end{align}

Compared with \cite{Liberzon2014}, where the same Lyapunov-like functions are employed to analyze stability of linear switched systems, we consider nonlinear switched systems.
Moreover, the switching conditions are different.

\begin{remark}
  Here, we explain the difference from the analysis of our previous work \cite{Kato2019ACC}.
  The functions in \eqref{eq:Lyapunov_function} are composed of two parts:
  The first part corresponds to the classical quadratic Lyapunov function and was used in \cite{Kato2019ACC} for stability analysis.
  Here, in addition, we have the second part related to the quantization error.
  If one employs the dynamic quantizer as explained in the previous section, then the quantization error is expected to converge to zero.
  Therefore, by adding the error term, one can utilize \eqref{eq:Lyapunov_function} as a Lyapunov function.
\end{remark}

The function $W_{\theta_k}(\xi_k,E_k)$ decreases under the nominal operation, whereas it increases under DoS attacks.
We now provide the convergence and divergence rates of this function depending on the occurrence of packet losses.
Let
\begin{align}
  \nu_0 &:= \max\{\varphi_0,\Lambda^2/M^2\}, \label{eq:nu0} \\
  \nu_1 &:= \max\{\varphi_1,\Lambda^2\}. \label{eq:nu1}
\end{align}
Then, the following lemma gives a local characterization of the switched Lyapunov-like function $W_{\theta_k}(\xi_k,E_k)$.
Now, in Lemma~\ref{lem:nonlinearity}, we choose $\gamma > 0$ sufficiently small such that $\delta < \varrho$.

\begin{lemma}
  Consider the nonlinear system \eqref{eq:plant} with \eqref{eq:control} as well as the dynamic quantizer \eqref{eq:xi} and \eqref{eq:E}.
  Suppose that Assumptions~\ref{assump:Lipschitz}--\ref{assump:initial_state} hold.
  Then, there exist $\omega_0 \in [\nu_0,1)$ and $\omega_1 \in [\nu_1,\infty)$ such that $\|\xi_k\|_\infty + E_k \le \delta$ implies
  \begin{align}
    W_{\theta_{k+1}}(\xi_{k+1},E_{k+1}) \le
    \begin{cases}
      \omega_{\theta_k} W_{\theta_k}(\xi_k,E_k) & \text{if $\theta_{k+1} = \theta_k$}, \\
      \mu_{\theta_k} \omega_{\theta_k} W_{\theta_k}(\xi_k,E_k) & \text{if $\theta_{k+1} \ne \theta_k$},
    \end{cases}
    \label{eq:evolution}
  \end{align}
  where $\mu_0$ and $\mu_1$ are as in \eqref{eq:mu}, and $\delta \in (0,\varrho)$ is given in Lemma~\ref{lem:nonlinearity}.
  \label{lem:Lyapunov_function}
\end{lemma}

\begin{IEEEproof}
  See Appendix~\ref{app:Lyapunov_function}.
\end{IEEEproof}

\begin{remark}
  The convergence and divergence rates $\omega_0$ and $\omega_1$ partly depend on the data rate of the communication channel.
  However, if the data rate is sufficiently large, then $\omega_0$ and $\omega_1$ converge to that of the infinite data rate case, which is determined only by the dynamics of the plant \eqref{eq:plant}.
  This property is the same as those of \cite{Liberzon2014}.
  In this case, we can recover our previous results presented in \cite{Kato2019ACC}.
  Furthermore, we have restricted ourselves to the case where the control input is reset to zero under DoS attacks.
  In this setting, it is not difficult to characterize the divergence rate under DoS attacks (see \eqref{eq:Lyapunov_inequality2}).
  We note that other control settings such as hold-input strategy \cite{DePersis-Tesi:TAC2015,DePersis-Tesi:sysconle2016} and output feedback \cite{Wakaiki2019} may be useful in practice.
  A similar analysis to this paper can be carried out although the characterization of a Lyapunov function as in \eqref{eq:evolution} becomes more complicated.
\end{remark}

\subsection{Stability Condition Under DoS Attacks}

Now, we are ready to state our main result.
Let $\kappa_D^* := \kappa_D + \kappa_F T$ and $\rho_D^* := \rho_D + \rho_F T$.
The following theorem extends the result of \cite{Kato2019ACC} to the case where quantization needs to be considered.

\begin{theorem}
  Consider the nonlinear networked control system \eqref{eq:plant} with the control input \eqref{eq:control}.
  Suppose that Assumptions~\ref{assump:Lipschitz}--\ref{assump:initial_state} hold.
  If
  \begin{align}
    \rho_F T \ln \mu_0 \mu_1 + (1 - \rho_D^*) \ln \nu_0 + \rho_D^* \ln \nu_1 < 0,
    \label{eq:stability}
  \end{align}
  then the origin is locally asymptotically stable.
  \label{thm:stability}
\end{theorem}

\begin{IEEEproof}
  Let $\chi(t)$ be the number of unsuccessful packet transmissions that occur in the time interval $[0,t]$.
  Using Assumptions~\ref{assump:frequency} and \ref{assump:duration}, we obtain
  \begin{align*}
    \chi(t) \le \frac{\kappa_D^* + \rho_D^* t}{T}.
  \end{align*}
  Since the quantizer does not saturate, i.e., \eqref{eq:quantizer_condition} holds, we have
  \begin{align*}
    \|x_k\|_\infty \le \|\xi_k\|_\infty + E_k.
  \end{align*}
  If $\|\xi_k\|_\infty + E_k \le \delta$ holds for all $k \in \Z_+$, then we obtain from Lemma~\ref{lem:Lyapunov_function} that
  \begin{align}
    W_{\theta_k}(\xi_k,E_k)
    &\le (\mu_0 \mu_1)^{N(t_k)} \omega_0^{k - \chi(t_k)} \omega_1^{\chi(t_k)} W_{\theta_0}(\xi_0,E_0) \nonumber \\
    &\le (\mu_0 \mu_1)^{\kappa_F + \rho_F t_k} \omega_0^{[- \kappa_D^* + (1 - \rho_D^*) t_k]/T} \nonumber \\
    &\quad{} \times \omega_1^{(\kappa_D^* + \rho_D^* t_k)/T} W_{\theta_0}(\xi_0,E_0) \nonumber \\
    &= c_W \omega^k W_{\theta_0}(\xi_0,E_0),
    \label{eq:W}
  \end{align}
  where $c_W := (\mu_0 \mu_1)^{\kappa_F} (\omega_1/\omega_0)^{\kappa_D^*/T}$ and $\omega := (\mu_0 \mu_1)^{\rho_F T} \omega_0^{1 - \rho_D^*} \omega_1^{\rho_D^*}$.
  From the choice of $\omega_0$ and $\omega_1$ respectively given by \eqref{eq:omega0} and \eqref{eq:omega1} in the proof of Lemma~\ref{lem:Lyapunov_function}, there always exists $\delta$ in Lemma~\ref{lem:Lyapunov_function} such that $\omega_0$ and $\omega_1$ are arbitrarily close to $\nu_0$ and $\nu_1$, respectively.
  The condition \eqref{eq:stability} thus implies that $\omega < 1$ holds in a certain small region, that is, small $\delta$.
  Next, we need to ensure that the quantization region is contained in such a small region.
  Since $\xi_0 = 0$, by choosing sufficiently small $E_0$, we have $\|\xi_k\|_\infty + E_k \le \delta$ for all $k \in \Z_+$.
  Therefore, the positive-definite function $W_{\theta_k}(\xi_k,E_k)$ converges to zero as $k \to \infty$, which implies asymptotic stability.
  Since the state lies in the quantization region at every sampling time under Assumption~\ref{assump:initial_state}, we can conclude the asymptotic stability of the origin.
\end{IEEEproof}

The stability condition \eqref{eq:stability} depends on the DoS parameters $\rho_F$ and $\rho_D$, which are characterized in Assumptions~\ref{assump:frequency} and \ref{assump:duration}.
The constants $\rho_F$ and $\rho_D$ give an upper bounds on the time-average of the number and the duration of DoS attacks, respectively.
Thus, the condition \eqref{eq:stability} requires that the average amount of DoS attacks is small enough.
In the absence of DoS attacks, the stability condition just requires that $\nu_0 < 1$, which is clearly satisfied from \eqref{eq:nu0}.
Notice that $\kappa_F$ and $\kappa_D$, which denote the initial energy for launching attacks, do not appear in the condition \eqref{eq:stability}.
However, these parameters are associated with the bound of the state trajectories and will be utilized in the analysis of the region of attraction in the next subsection.

\begin{remark} 
  Here, we explain the comparison with the existing results on networked control under DoS attacks.
  The authors of \cite{DePersis-Tesi:TAC2015} investigate input-to-state stability for linear plants with respect to disturbances under more restrictive class of DoS attacks.
  The remainder term of linearization as well as measurement errors due to quantization can be seen as a special case of disturbances.
  However, the nonlinear term has the property that its effects vanish at the origin.
  Also, quantization errors converge to zero as we employ a dynamic quantizer.
  By these properties, we can use Assumptions~\ref{assump:frequency} and \ref{assump:duration} in DoS models instead of more restrictive class (see also Remark~\ref{rem:dos}).
  The stability condition \eqref{eq:stability} is similar to that of \cite{DePersis-Tesi:TAC2015} (see also \cite{DePersis-Tesi:sysconle2016} for the nonlinear systems case).
  As our focus is on a linearization approach, we can recover the global stability result for linear systems by ignoring the nonlinear parts in \eqref{eq:linearization}.
  Compared with \cite{DePersis-Tesi:sysconle2016}, we explored local stability of the nonlinear system \eqref{eq:plant} particularly in the linearization framework.
  As we discuss in the next subsection, the local stability point of view is important when DoS attacks are addressed in stabilization problems.
\end{remark}

\begin{remark}
  The dynamic quantizer proposed in this paper is resilient in the sense that it does not saturate even under DoS attacks.
  The above theorem can also be seen as an extension of the work \cite{Liberzon-Hespanha:TAC2005}, where the effects of packet losses are not considered.
  Furthermore, we take into account the unstable dynamics induced by DoS attacks.
  The condition in the above theorem indicates the allowable average frequency and duration of such attacks to preserve local stability of the nonlinear system.
  Notice that if the data rate is appropriately large, then we have $\mu_0 = \lambda_\mathrm{max}(P_1)/\lambda_\mathrm{min}(P_0)$, $\mu_1 = \lambda_\mathrm{max}(P_0)/\lambda_\mathrm{min}(P_1)$, $\nu_0 = \varphi_0$, and $\nu_1 = \mu_1$ in \eqref{eq:mu0}--\eqref{eq:nu1}.
  These parameters are consistent with those of the stability condition in the case of the infinite data rate which is presented in \cite{Kato2019ACC}.
  As mentioned in Remark~\ref{rem:data_rate}, it is difficult to find an appropriate coordinate transformation applied in the quantization process as in the linear systems case.
  In particular, the choice of the coordinate transformation affects the estimate of the reachable set, which is associated with the zooming-in/out procedure.
  Thus, investigating more explicit relationship between the limitation of quantized control and the tolerance of DoS attacks for nonlinear systems is left to future work.
\end{remark}

\subsection{Convergence Condition on Initial States}

In the previous part of this section, we derived a local stability condition.
Due to linearization, we need to keep the state within a small region around the equilibrium even in the presence of packet losses.
Otherwise, the state cannot converge to the equilibrium point.
In particular, we need to set the initial condition so that the inequality \eqref{eq:evolution} is satisfied.
This is because that inequality may not be valid when $\|\xi_k\|_\infty + E_k > \delta$.
The following theorem provides a condition on $E_0$ that guarantees the state trajectory to stay inside the stability region at all times and eventually converge to the origin.

\begin{theorem}
  Consider the nonlinear networked control system \eqref{eq:plant} with the control input \eqref{eq:control}.
  Suppose that Assumptions~\ref{assump:Lipschitz}--\ref{assump:initial_state} hold.
  Let $\omega_0$, $\omega_1$, and $\delta$ be taken from Lemma~\ref{lem:Lyapunov_function}.
  Also, suppose that \eqref{eq:stability} holds.
  If we choose $E_0$ to satisfy
  \begin{align}
    E_0 < (\mu_0 \mu_1)^{-\kappa_F/2} \left( \frac{\omega_0}{\omega_1} \right)^{\kappa_D^*/(2T)} \delta^*,
    \label{eq:convergence}
  \end{align}
  where $\delta^* := \delta \sqrt{\alpha/\beta}$, then the state trajectory $x(t)$ remains within the set $\{x \in \R^n : \|x\|_\infty < \delta\}$ for all $t \ge 0$ and moreover achieves $\lim_{t \to \infty} \|x(t)\|_\infty = 0$.
  \label{thm:convergence}
\end{theorem}

\begin{IEEEproof}
  Recall from \eqref{eq:W} that, by Lemma~\ref{lem:Lyapunov_function}, if $\|\xi_k\|_\infty + E_k < \delta$, then
  \begin{align*}
    W_{\theta_k}(\xi_k,E_k) \le c_W \omega^k W_{\theta_0}(\xi_0,E_0).
  \end{align*}
  Under the condition \eqref{eq:stability}, it holds that $\omega < 1$, and hence, we obtain
  \begin{align*}
    \|\xi_k\|_\infty + E_k \le \sqrt{\frac{\beta}{\alpha}} c_W^{1/2} (\|\xi_0\|_\infty + E_0),
  \end{align*}
  where we have used the inequalities \eqref{eq:alpha_beta}.
  Since $\xi_0 = 0$, the above inequality becomes
  \begin{align*}
    \|\xi_k\|_\infty + E_k \le \sqrt{\frac{\beta}{\alpha}} c_W^{1/2} E_0.
  \end{align*}
  Note that \eqref{eq:convergence} can be written as
  \begin{align*}
    E_0 < \sqrt{\frac{\alpha}{\beta}} c_W^{-1/2} \delta.
  \end{align*}
  Thus, it follows $\|\xi_k\|_\infty + E_k < \delta$ for all $k \in \Z_+$.
  When the state lies within the quantization region at time $t_k$, we have $\|x_k\|_\infty \le \|\xi_k\|_\infty + E_k$.
  Since $W_{\theta_k}(\xi_k,E_k)$ converges to zero, we can guarantee that the state $x(t)$ approaches the origin.
\end{IEEEproof}

\begin{remark}
  The result in Theorem~\ref{thm:convergence} is important in the sense that the condition \eqref{eq:convergence} may not hold while the stability condition \eqref{eq:stability} holds.
  Such a case occurs when the DoS parameters $\kappa_F$ and $\kappa_D$ are large.
  This property is not discussed in \cite{DePersis-Tesi:sysconle2016} since the authors consider global stability.
  In practice, it is important to focus on the effects of DoS attacks to the region of attraction.
  The above theorem provides a quantitative condition under which the state trajectory can remain within the nominal region of attraction arising due to linearization.
  Here, we emphasize that a certain level of DoS attacks makes the state go outside the region of attraction, possibly leading to an unstable behavior.
  Therefore, from the viewpoint of local stability, the initial state should be close enough to the equilibrium point if DoS attacks are present.
\end{remark}

\section{Simulation Example}
\label{sec:simulation}

Here, we demonstrate the efficacy of our main results through a simulation example.

Consider the Li{\'e}nard system
\begin{align*}
  \ddot{z}(t) - (1 - 3az^2(t) - 5bz^4(t)) \dot{z}(t) + z(t) = u(t),
\end{align*}
where $a = 1/3$ and $b = 1/50$.
Choosing the state as
\begin{align*}
  x(t) =
  \begin{bmatrix}
    x_1(t) \\ x_2(t)
  \end{bmatrix}
  =
  \begin{bmatrix}
    z(t) \\ \dot{z}(t) - \int_0^{z(t)} (1 + 3aw^2 - 5bw^4)\,\diff w
  \end{bmatrix},
\end{align*}
we obtain the state equation
\begin{align*}
  \begin{bmatrix}
    \dot{x}_1(t) \\
    \dot{x}_2(t) \\
  \end{bmatrix}
  =
  \begin{bmatrix}
    x_2(t) + x_1(t) + ax_1^3(t) - bx_1^5(t) \\
    -x_1(t) + u(t)
  \end{bmatrix}.
\end{align*}
The right-hand side of the above equation is locally Lipschitz with $L = 10$, satisfying Assumption~\ref{assump:Lipschitz}.
Also, we choose the sampling period as $T = 0.1$ and the number of quantization levels as $M = 6$.
The uncontrolled system has an unstable equilibrium point at the origin and exhibits a stable limit cycle.

\begin{figure}
  \centering
  \includegraphics[scale=0.38]{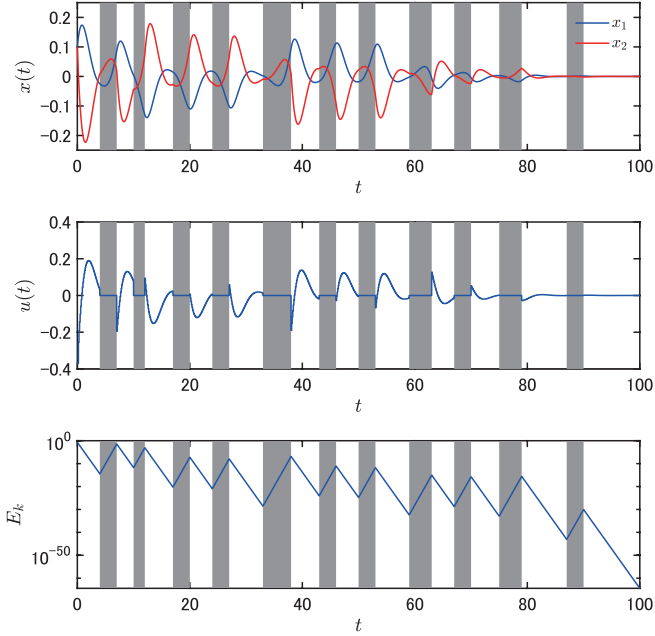}
  \caption{Trajectories of system state, input, and size of the quantization range}
  \label{fig:simulation}
\end{figure}

To stabilize the origin, we consider our linearization-based quantized control framework.
Specifically, we set the feedback gain to $K = [-1.81\;{-1.90}]$, which is obtained by using the LQR method on the linearized system.
The simulation result is presented in Fig.~\ref{fig:simulation}, where the initial state is set to $x_0 = [0.1\;0.1]^\T$.
In the figure, the shaded parts represent the DoS attack intervals.
The bottom figure shows the changes in the radius $E_k$ of the quantizer.
One can observe that saturation is avoided by expanding the quantization region when DoS is present.
From the simulation result, we can see that the state $x(t)$ converges to the origin under DoS attacks.

\begin{figure}
  \centering
  \includegraphics[scale=0.4]{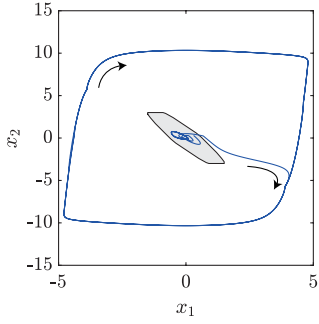}
  \caption{State trajectory that leaves the region of attraction and approaches a limit cycle. The shaded area in gray represents the nominal region of attraction.}
  \label{fig:unstable_phase_portrait}
\end{figure}

Then, we explain the importance to consider nonlinear systems in the context of DoS attacks.
Due to linearization, if the initial state is located far from the equilibrium, then the state trajectory from that position leaves the region of attraction and converges to a limit cycle trajectory.
In such cases, the state is unable to go to the origin by the linearization-based control even after the communication recovers.
This fact can be observed in Fig.~\ref{fig:unstable_phase_portrait}, where the initial state is set to $x_0 = [0.3\;0.3]^\T$ and DoS attacks are kept the same as above.
Here, the shaded area in gray represents the nominal region of attraction.
This area is numerically obtained by finding states such that trajectories starting from there without the effects of quantization and DoS attacks converge to the origin.
Note that the Lipschitz continuity of $f$ with Lipschitz constant $L = 10$ is preserved in this region, that is, the region $\mathcal{D}$ in Assumption~\ref{assump:Lipschitz} is larger than the region of attraction.
Also, notice that in the simulation in Fig.~\ref{fig:unstable_phase_portrait}, the initial state is within this region.
Thus, the undesired unstable phenomenon is due to the nonlinearity of the plant induced by the DoS attacks.

\begin{figure}
  \centering
  \includegraphics[scale=0.4]{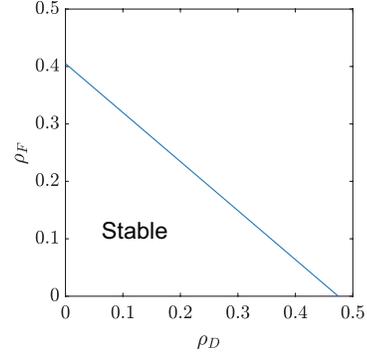}
  \caption{Allowable DoS attack level. At the lower left area, stability of the origin is preserved.}
  \label{fig:dos_parameters}
\end{figure}

Here, we provide some discussion on the theoretical results in the previous section.
The stability condition derived in Theorem~\ref{thm:stability} is presented in Fig.~\ref{fig:dos_parameters}.
Under the DoS parameters at the lower left area, the stability of the origin is preserved.
Thus, if the initial state is very close to the origin, the trajectory can converge to the origin even in the presence of DoS attacks.
However, we need to emphasize that the region of attraction is affected by the strength of DoS attacks.
In Theorem~\ref{thm:convergence}, the theoretical value of $\delta$ is $\delta = 1.94 \times 10^{-7}$.
In the presence of DoS attacks, the estimated region of attraction becomes much smaller.
This theoretical result is indeed quite conservative, and some numerical methods can be used to gain more precise estimate of the region of attraction as above.
Theoretical study on the relation between the region of attraction and DoS attacks is an important direction of future work.
For example, there are vulnerable positions in the state space from which it is easy for the attacker to make the state leave the region of attraction.

\section{Conclusion}
\label{sec:conclusion}

In this paper, we have considered a quantized stabilization problem of nonlinear networked control systems under DoS attacks.
Our proposed control strategy is based on the linearization framework used together with a resilient dynamic quantizer which does not saturate in the presence of packet losses.
A sufficient condition for stability and an estimate of the region of attraction have been derived, characterizing tolerable frequency and duration of DoS attacks.
The simulation example demonstrates our results.
Future research includes synchronization of nonlinear multi-agent systems under DoS attacks, where information is exchanged among spatially distributed agents.
Furthermore, resilient control against DoS attacks by using prediction of lost measurements is another interesting direction.

\appendices

\section{Proof of Lemma~\ref{lem:inter-sample}}
\label{app:inter-sample}

For $\bar{x} \in \R^n$, the solution $\phi(t,\bar{x},\bar{u})$ to \eqref{eq:plant} can be written
\begin{align*}
  \phi(t,\bar{x},\bar{u}) = \bar{x} + \int_0^t [A \phi(s,\bar{x},\bar{u}) + B \bar{u} + g(\phi(s,\bar{x},\bar{u}),\bar{u})] \,\diff s
\end{align*}
for $t \in [0,T)$.
From Taylor's theorem, we have
\begin{align*}
  \lim_{(x,u) \to (0,0)} \frac{\|g(x,u)\|_\infty}{\sqrt{\|x\|_\infty^2 + \|u\|_\infty^2}} = 0.
\end{align*}
It follows that there exists a positive constant $d^\prime > 0$ such that
\begin{align}
  \sqrt{\|x\|_\infty^2 + \|u\|_\infty^2} < d^\prime \implies \|g(x,u)\|_\infty \le \|x\|_\infty + \|u\|_\infty.
  \label{eq:d'}
\end{align}
Now, suppose that $\sqrt{\|\phi(t,\bar{x},\bar{u})\|_\infty^2 + \|\bar{u}\|_\infty^2} < d^\prime$ holds for all $t \in [0,T)$.
Then, substituting $\bar{u} = (1 - \theta) K \bar{x}$ yields
\begin{align*}
  \|\phi(t,\bar{x},\bar{u})\|_\infty
  &\le [1 + (1 - \theta) T (\|BK\|_\infty + \|K\|_\infty)] \|\bar{x}\|_\infty \\
  &\quad{} + \int_0^t (\|A\|_\infty + 1) \|\phi(s,\bar{x},\bar{u})\|_\infty \,\diff s.
\end{align*}
Applying Gronwall's inequality, we obtain
\begin{align*}
  &\|\phi(t,\bar{x},\bar{u})\|_\infty \\
  &\le [1 + (1 - \theta) T (\|BK\|_\infty + \|K\|_\infty)] \|\bar{x}\|_\infty \e^{\int_0^t (\|A\|_\infty + 1) \,\diff s} \\
  &\le [1 + (1 - \theta) T (\|BK\|_\infty + \|K\|_\infty)] \e^{T (\|A\|_\infty + 1)} \|\bar{x}\|_\infty.
\end{align*}
Let $d := d^\prime/\sqrt{c_0^2 + \|K\|_\infty^2}$.
In this case, we observe that $\sqrt{\|\phi(t,\bar{x},\bar{u})\|_\infty^2 + \|\bar{u}\|_\infty^2} < d^\prime$ is satisfied whenever $\|\bar{x}\|_\infty + \|\bar{u}\|_\infty < d$.
Thus, by \eqref{eq:d'}, we obtain the desired result.
\hfill \IEEEQEDhere

\section{Proof of Lemma~\ref{lem:nonlinearity}}
\label{app:nonlinearity}

It can be seen that for any $\gamma > 0$, there exists a constant $\delta^\prime > 0$ such that $\sqrt{\|x\|_\infty^2 + \|u\|_\infty^2} < \delta^\prime$ implies $\|g(x,u)\|_\infty \le \gamma (\|x\|_\infty + \|u\|_\infty)$.
With the scalar $d$ given in Lemma~\ref{lem:inter-sample}, define $\delta := \min \{ d,\delta^\prime/\sqrt{c_0^2 + \|K\|_\infty^2}\}$.
Whenever $\|\bar{x}\|_\infty + \|\bar{u}\|_\infty < \delta$, we have $\sqrt{\|\phi(t,\bar{x},\bar{u})\|_\infty^2 + \|\bar{u}\|_\infty^2} \le \delta$ for all $t \in [0,T)$.
It thus follows
\begin{align*}
  &\|\tilde{g}(\bar{x},\bar{u})\|_\infty \\
  &\le \gamma \e^{T \|A\|_\infty} \int_0^T [\|\phi(t,\bar{x},\bar{u})\|_\infty + (1 - \theta) \|K\|_\infty \|\bar{x}\|_\infty] \,\diff s \\
  &\le \gamma T \e^{T \|A\|_\infty} [c_\theta \|\bar{x}\|_\infty + (1 - \theta) \|K\|_\infty \|\bar{x}\|_\infty] \le \gamma_{\theta} \|\bar{x}\|_\infty,
\end{align*}
where the second inequality follows from Lemma~\ref{lem:inter-sample}.
This completes the proof.
\hfill \IEEEQEDhere

\section{Proof of Lemma~\ref{lem:Lyapunov_function}}
\label{app:Lyapunov_function}

We consider the two cases of $\theta_{k+1} = \theta_k = 0$ and $\theta_{k+1} = \theta_k = 1$, separately.
At first, we consider the case where $\theta_{k+1} = \theta_k = 0$.
It follows from \eqref{eq:xi} that
\begin{align*}
  \xi_{k+1} = F(q_k,Kq_k) = \Phi_0 q_k + h_0(q_k)
\end{align*}
with $\Phi_0 := \tilde{A} + \tilde{B} K$ and $h_0(q_k) := \tilde{g}(q_k,Kq_k)$.
Let us define the positive-definite function $V_0(\xi) := \xi^\T P_0 \xi$ for $\xi \in \R^n$.
Then, this function satisfies
\begin{align*}
  V_0(\xi_{k+1}) &= q_k^\T \Phi_0^\T P_0 \Phi_0 q_k + 2 h_0^\T(q_k) P_0 \Phi_0 q_k \\
                 &\quad{} + h_0^\T(q_k) P_0 h_0(q_k) \\
                 &\le \varphi_0 q_k^\T P_0 q_k + 2 \|P_0 \Phi_0\|_\infty \|q_k\|_\infty \|h_0(q_k)\|_\infty \\
                 &\quad{} + \|P_0\|_\infty \|h_0(q_k)\|_\infty^2,
\end{align*}
where we have used \eqref{eq:Lyapunov_inequality1} in the inequality.
By applying Lemma~\ref{lem:nonlinearity}, it holds that if $\|\xi_k\|_\infty + E_k < \delta$, which yields $\|q_k\|_\infty < \delta$, then $\|h_0(q_k)\|_\infty \le \gamma_0 \|q_k\|_\infty$.
Thus, we have
\begin{align*}
  V_0(\xi_{k+1}) \le \hat{\varphi}_0 V_0(q_k).
\end{align*}
where $\hat{\varphi}_0 := \varphi_0 + (2 \gamma_0 \|P_0 \Phi_0\|_\infty + \gamma_0^2 \|P_0\|_\infty)/\lambda_\mathrm{min}(P_0)$, and $\lambda_\mathrm{min}(\cdot)$ represents the minimum eigenvalue of a matrix.
Here, we define $\zeta_k := q_k - \xi_k$.
Then, it satisfies $\|\zeta_k\|_\infty < (M - 1)/M$.
Moreover, we obtain
\begin{align*}
  V_0(q_k) &= \xi_k^\T P_0 \xi_k + 2 \zeta_k^\T P_0 \xi_k + \zeta_k^\T P_0 \zeta_k \\
           &\le \xi_k^\T P_0 \xi_k + 2 \|P_0\|_\infty \|\xi_k\|_\infty \|\zeta_k\|_\infty + \|P_0\|_\infty \|\zeta_k\|_\infty^2.
\end{align*}
From Young's inequality, for any positive number $\varepsilon > 0$, it holds
\begin{align*}
  2 \|\xi_k\|_\infty \|\zeta_k\|_\infty \le \frac{1}{\varepsilon} \|\xi_k\|_\infty^2 + \varepsilon \|\zeta_k\|_\infty^2.
\end{align*}
By using this, the above inequality becomes
\begin{align*}
  V_0(q_k) \le \tilde{\varphi}_0 V_0(\xi_k) + \vartheta \|\zeta_k\|_\infty^2
\end{align*}
with the constants $\tilde{\varphi}_0 := \hat{\varphi}_0 + \|P_0\|_\infty/(\varepsilon \lambda_\mathrm{min}(P_0))$ and $\vartheta := (1 + \varepsilon) \|P_0\|_\infty$.
Note that one can always choose a large $\varepsilon$ to guarantee $\tilde{\varphi}_0 < 1$ since $\hat{\varphi}_0 < 1$ by hypothesis.
Finally, it follows that
\begin{align*}
  V_0(\xi_{k+1}) \le \tilde{\varphi}_0 V_0(\xi_k) + \vartheta \left( \frac{M-1}{M} \right)^2 E_k^2.
\end{align*}
Therefore, from \eqref{eq:E} and \eqref{eq:Lyapunov_function}, we obtain
\begin{align}
  &W_0(\xi_{k+1},E_{k+1}) \nonumber \\
  &= V_0(\xi_{k+1}) + \vartheta \left( \frac{M-1}{M} \right)^2 E_k^2 + \eta_0 \frac{\Lambda^2}{M^2} E_k^2 \nonumber \\
  &= \tilde{\varphi}_0 V_0(\xi_k) + \vartheta \left( \frac{M-1}{M} \right)^2 E_k^2 + \eta_0 \frac{\Lambda^2}{M^2} E_k^2 \nonumber \\
  &\le \omega_0 W_0(\xi_k,E_k),
  \label{eq:W0}
\end{align}
where
\begin{align}
  \omega_0 := \max \left\{ \tilde{\varphi}_0, \frac{\vartheta}{\eta_0} \left( \frac{M-1}{M} \right)^2 + \frac{\Lambda^2}{M^2} \right\}.
  \label{eq:omega0}
\end{align}
By Assumption~\ref{assump:data_rate}, there always exists $\eta_0 > 0$ such that $\omega_0 < 1$.

Next, consider the case where $\theta_{k+1} = \theta_k = 1$.
If this is the case, the quantizer \eqref{eq:xi} can be written by
\begin{align*}
  \xi_{k+1} = f(\xi_k,0) = \Phi_1 \xi_k + h_1(\xi_k),
\end{align*}
where $\Phi_1 := \tilde{A}$ and $h_1(\xi_k) := \tilde{g}(\xi_k,0)$.
We also define $V_1(x) := x^\T P_1 x$ for all $x \in \R^n$.
From \eqref{eq:Lyapunov_inequality2},
\begin{align*}
  V_1(\xi_{k+1}) &\le \varphi_1 \xi_k^\T P_1 \xi_k + 2 \|P_1 \Phi_1\|_\infty \|\xi_k\|_\infty \|h_1(\xi_k)\|_\infty \\
                 &\quad{} + \|P_1\|_\infty \|h_1(\xi_k)\|_\infty^2.
\end{align*}
It then follows from Lemma~\ref{lem:nonlinearity} that if $\|\xi_k\|_\infty + E_k < \delta$,
\begin{align*}
  V_1(\xi_{k+1}) \le \hat{\varphi}_1 V_1(\xi_k),
\end{align*}
where $\hat{\varphi}_1 := \varphi_1 + (2 \gamma_1 \|P_1 \Phi_1\|_\infty + \gamma_1^2 \|P_1\|_\infty)/\lambda_\mathrm{min}(P_1)$.
We therefore obtain from \eqref{eq:E} that
\begin{align}
  W_1(\xi_{k+1},E_{k+1}) &= V_1(\xi_{k+1}) + \eta_1 E_{k+1}^2 \nonumber \\
                         &\le \hat{\varphi}_1 \xi_k^\T P_1 \xi_k + \Lambda^2 \eta_1 E_k^2 \nonumber \\
                         &\le \omega_1 W_1(\xi_k,E_k), \label{eq:W1}
\end{align}
where
\begin{align}
  \omega_1 := \max \{\hat{\varphi}_1, \Lambda^2\} > 1.
  \label{eq:omega1}
\end{align}

Therefore, in \eqref{eq:W0} and \eqref{eq:W1}, we obtained the desired result \eqref{eq:evolution} for the case where $\theta_{k+1} = \theta_k$.
The relation for $\theta_{k+1} \ne \theta_k$ can be found by further applying the inequalities in \eqref{eq:mu} to \eqref{eq:W0} and \eqref{eq:W1}.
The proof is now complete.
\hfill \IEEEQEDhere

\bibliographystyle{IEEEtran}
\bibliography{IEEEtran}

\end{document}